\documentclass[conference,compsoc]{IEEEtran}
\ifCLASSOPTIONcompsoc
  \usepackage[nocompress]{cite}
\else
  \usepackage{cite}
\fi
\ifCLASSINFOpdf
\else
\fi
\usepackage{cite}
\usepackage{amsmath,amssymb,amsfonts}
\usepackage{algorithmic}
\usepackage{graphicx}
\usepackage{multirow}
\usepackage{textcomp}
\usepackage{float}
\usepackage{subcaption}
\usepackage{arydshln}
\usepackage{tikz}
\usepackage{subcaption}
\usepackage{subfiles}
\usepackage [english]{babel}
\usepackage [autostyle, english = american]{csquotes}
\MakeOuterQuote{"}
\usepackage{fancyhdr}
\usepackage{url}
\def\BibTeX{{\rm B\kern-.05em{\sc i\kern-.025em b}\kern-.08em
    T\kern-.1667em\lower.7ex\hbox{E}\kern-.125emX}}

\begin{document}

\title{Nuclei Instance Segmentation of Cryosectioned H\&E Stained Histological Images using Triple U-Net Architecture }

\author{
	\IEEEauthorblockN{Chowdhury Nur E Alam Siddiqi\IEEEauthorrefmark{1},
		Fardifa Fathmiul Alam\IEEEauthorrefmark{1},
		Zarif Ahmed\IEEEauthorrefmark{1},
		Tasnim Ahmed\IEEEauthorrefmark{2} and\\
		Tareque Mohmud Chowdhury\IEEEauthorrefmark{3}
	}
	\IEEEauthorblockA{
		\IEEEauthorrefmark{1}\IEEEauthorrefmark{1}\IEEEauthorrefmark{1}\IEEEauthorrefmark{2}\IEEEauthorrefmark{3}
		Department of Computer Science and Engineering, Islamic University of Technology, Dhaka, Bangladesh\\
		}
	\IEEEauthorblockA{
		\{\IEEEauthorrefmark{1}{nurealam, \IEEEauthorrefmark{1}fardifafathmiul, \IEEEauthorrefmark{1}zarifahmed, \IEEEauthorrefmark{2}tasnimahmed, \IEEEauthorrefmark{3}tareque}\}
		@iut-dhaka.edu
	}
}

\renewcommand{\footnoterule}
    {\noindent\smash{\rule[3pt]{\columnwidth}{0.4pt}}}


% compsoc conferences compared to traditional conferences), use this
%Telephone: (800) 555--1212, Fax: (888) 555--1212}


\maketitle

\thispagestyle{fancy}
\pagenumbering{gobble}
\renewcommand{\headrulewidth}{0pt}
% \fancyhead[R]{}
% \fancyfoot[C]{978-1-6654-3902-2/21/\$31.00 ©2021 IEEE}


\begin{abstract}

Nuclei instance segmentation is crucial in oncological diagnosis and cancer pathology research. H\&E stained images are commonly used for medical diagnosis, but pre-processing is necessary before using them for image processing tasks. Two principal pre-processing methods are formalin-fixed paraffin-embedded samples (FFPE) and frozen tissue samples (FS). While FFPE is widely used, it is time-consuming, while FS samples can be processed quickly. Analyzing H\&E stained images derived from fast sample preparation, staining, and scanning can pose difficulties due to the swift process, which can result in the degradation of image quality. This paper proposes a method that leverages the unique optical characteristics of H\&E stained images. A three-branch U-Net architecture has been implemented, where each branch contributes to the final segmentation results. The process includes applying watershed algorithm to separate overlapping regions and enhance accuracy. The Triple U-Net architecture comprises an RGB branch, a Hematoxylin branch, and a Segmentation branch. This study focuses on a novel dataset named CryoNuSeg, the first-ever FS Dataset featuring 30 FS-sectioned images of 10 human organs. The results obtained through robust experiments outperform the state-of-the-art results across various metrics. The benchmark score for this dataset is AJI 52.5 and PQ 47.7, achieved through the implementation of U-Net Architecture. However, the proposed Triple U-Net architecture achieves an AJI score of 67.41 and PQ of 50.56. The proposed architecture improves more on AJI than other evaluation metrics, which further justifies the superiority of the Triple U-Net architecture over the baseline U-Net model, as AJI is a more strict evaluation metric since it penalizes wrong predictions more considerably than it awards correct predictions. The use of the three-branch U-Net model, followed by watershed post-processing, significantly surpasses the benchmark scores, showing substantial improvement in the AJI score.

\end{abstract}



% \fi
\IEEEpeerreviewmaketitle

\section{Introduction}
\label{sec:intro}
Cancer is a leading cause of global mortality, claiming nearly 10 million lives in 2020 alone \cite{ferlay2020gco}. Timely implementation of evidence-based prevention measures can prevent 30 to 50 percent of cancer cases. Early cancer detection is crucial, increasing the chances of successful treatment and recovery. Accurate and swift cancer diagnosis is vital during surgical procedures to guide immediate treatment decisions.

Nucleus holds the majority of the cell's genetic materials and has been a primary focus in cancer research, often referred to as the "gene-centric view" of cancer. The identification of oncogenes, tumor-suppressor genes, and the understanding of multiple cell mutations have become standard prerequisites for comprehending cancer initiation and progression.

In the context of cancer diagnosis, the segmentation and extraction of nuclei regions from cell tissue images hold immense significance. This process, known as nuclei segmentation, is primarily performed on Hematoxylin and Eosin (H\&E)-stained images, the most widely used staining technique in pathological diagnosis \cite{titford2005long}. The Hematoxylin stain imparts a purplish-blue color to nuclei. In contrast, the Eosin stain imparts a pink color to the extracellular matrix and cytoplasm, resulting in distinct color variations in different parts of the tissue cell. Detailed analysis of H\&E-stained tissue sections provides critical insights into each cell's functional situation, making it the "gold standard" in diagnosing various types of cancer \cite{chan2014wonderful}. Assessing cell images stained with H\&E relies on essential criteria like the morphology, shape, type, count, and density of nuclei. Automating the extraction of these characteristics using digital techniques necessitates the segmentation of nuclei.

Being able to accurately identify tumor cells from nuclei segmentation quickly is essential for cancer detection. Several image segmentation techniques have been developed for traditional segmentation tasks. Most common methods fail to accomplish the accurate segmentation of medical images, especially nuclei segmentation, as segmenting the nucleus requires the production of masks that have exact boundaries around each nucleus and should differentiate overlapping regions. This paper proposes the Triple U-Net architecture to perform instance segmentation of cell nuclei in the CryoNuSeg dataset - the first fully annotated FS-derived cryosectioned and H\&E stained nuclei dataset. The proposed algorithm uses the Triple U-Net model instead of the baseline U-Net model. The motive is to present a state-of-the-art architecture for high-precision nuclear segmentation, facilitating quick cancer diagnosis with greater precision.

\section{Related Works}
\label{sec:related-works}
In the field of computer vision, image segmentation is a fundamental process that involves dividing the pixels of an image into distinct areas, each labeled as a particular part or segment. The segmentation process is essential as it enables the extraction of regions of interest, facilitating in-depth analysis and understanding of these regions. As computing power is increasing, nuclei segmentation has become a superior methodology in the diagnosis of tumors and cancers. Recent advancements in deep learning have given rise to some powerful image segmentation architectures that are able to perform both instance and semantic segmentation with so much accuracy that even cell nuclei can be segmented. Two main approaches for image segmentation have been explored by researchers: Classical Non-AI-Based Approaches and AI-Based Approaches.

\subsection{Classical Non-AI Based Approaches}
Classical Non-AI Based Approaches encompass a range of algorithms that have been developed and studied extensively over the years. These methods include thresholding \cite{al-amri2010image}, histogram specification \cite{thomas2011histogram}, region growing \cite{gomez2007image} , k-means clustering \cite{jin2011kmeans}, watersheds \cite{roerdink2001watershed}, active contours \cite{pluempitiwiriyawej2005stacs}, graph cuts \cite{aplimat2015proceedings}, conditional and Markov random fields \cite{wu2019imagelabeling}, and sparsity-based \cite{sparsityclassification2011} methods. Although these techniques have been valuable, their accuracy has been limited, and they have faced challenges in multi-object detection.

\subsection{AI Based Approaches}

\subsubsection{Convolutional Neural Network}
With the advent of image recognition tasks, the need for a better image-suitable neural network was required. Traditional Feed Forward neural networks fell short in achieving satisfactory accuracy for such tasks. Consequently, the Convolutional Neural Network (CNN) emerged as a result of extensive research efforts \cite{oshea2015introduction}. CNNs, as the name implies, leverage the Convolution Operation on input data. This operation plays a pivotal role in preserving spatial information, contextual details, and temporal relationships within the image data. Consequently, CNNs are well-suited for processing higher-dimensional data, particularly images. Their ability to retain crucial spatial and contextual features makes them ideal for image-related applications.

Researchers have also explored combining CNNs with other algorithms to enhance segmentation tasks. One such example is the integration of CNNs with the Connected Component algorithm (CC) for the segmentation of Scanning Electron Microscopy (SEM) images \cite{cnnsegmentation2011}. This hybrid approach has shown promising results in accurately segmenting SEM images, further demonstrating the versatility and potential of CNNs in medical image analysis.

\subsubsection{Fully Convolutional Neural Network}
The traditional Convolutional Neural Network (CNN) architecture employs convolution and pooling operations to down-sample the input image and learn a condensed feature map. This is done mainly to make the whole algorithm computationally feasible for implementation. This downsampling is crucial for computational feasibility, especially in simple image classification tasks where identifying the object's presence suffices without precise localization information. However, for the segmentation task, localization information is necessary to locate the object in the image, but the downsampling does not conserve the localization information.

To mitigate this problem and make image segmentation feasible without making it computationally expensive, a new design architecture was proposed called the "Fully Convolutional Neural Network" \cite{long2015fully}. FCN introduces an Upsampling component following the traditional CNN's downsampling phase, enabling the incorporation of location information along with contextual details for semantic segmentation. Several research works have been done using FCN that produced impressive semantic segmentation results. For instance, \cite{chen2018deeplab} and \cite{yang2019deeperlab} have introduced an image parser that performs whole image parsing using FCN, addressing both semantic and instance segmentation tasks using a single-shot bottom-up approach.

\subsubsection{U-Net}
U-Net \cite{ronneberger2015unet} has emerged as one of the most widely adopted models for biomedical image segmentation, utilizing the architectural foundation of Fully Convolutional Neural Networks (FCNs). The U-Net model is characterized by its encoder-decoder structure, which facilitates efficient feature extraction and up-sampling for precise segmentation outcomes.

The encoder component of U-Net performs down-sampling of the input image, akin to conventional CNN architectures, generating a condensed feature map. This condensed feature map is then fed into the decoder, which undertakes the crucial up-sampling process to restore the image to its original resolution. Notably, U-Net employs 3x3 filters and 2x2 pooling layers for the entire down-sampling procedure, and each layer increases the feature maps two-fold.

To achieve the up-sampling, U-Net relies on the fundamental principle of transpose convolution, which performs the inverse of the convolution operation and enlarges the feature maps. Moreover, U-Net's distinctiveness lies in the concatenation of the encoder's feature maps with those of the decoder. This concatenation strategy enhances the network's understanding and representation capacity during the training process, leading to improved segmentation performance.

The activation function adopted in U-Net is the Rectified Linear Unit (ReLU) \cite{agarap2018deeprelu}, a widely acclaimed choice in neural network architectures. ReLU has demonstrated its efficiency in multiple research papers and contributes to the model's ability to capture complex patterns and features in biomedical images effectively.

\subsubsection{Double U-Net}
This \cite{jha2020doubleunet} was introduced as an improvement of the U-Net architecture, enhancing its performance on various segmentation tasks, specifically in the medical domain, such as colonoscopy and dermoscopy datasets. It is a combination of two U-Net architectures stacked on top of each other—one of them utilizes a pretrained VGG-19 encoder, and another U-Net is added for capturing greater semantic features more efficiently.

\subsubsection{Triple U-Net}

This network \cite{ZHAO2020101786} presents a novel architecture comprising three U-Net branches, each serving distinct goals in the context of nuclei segmentation from Whole Slide Images (WSI). In particular, the RGB component is dedicated to capturing the fundamental characteristics of the segmentation task in its raw form. On the contrary, the H branch employs a feature extraction method that is sensitive to Hematoxylin, taking a distinct approach., enabling precise detection of nuclei edges and enhancing the network's ability to differentiate overlapping regions in the images. This Hematoxylin-aware contour feature extraction is especially beneficial for obtaining more accurate segmentation masks of cryosectioned samples of whole slide images.

An essential aspect of the proposed architecture is the stability of the Hematoxylin module concerning the inconsistent color of Hematoxylin and Eosin (H\&E) stains. Consequently, the authors opted not to apply normalization, as doing so could potentially lead to unnecessary information loss. The segmentation branch efficiently combines the raw features from the RGB branch with the Hematoxylin-sensitive contour features from the H branch to make predictions for the final segmentation outcomes.

The authors' introduction of a Hematoxylin-aware Triple U-Net signifies an innovative and original approach, aiming to exploit the specific properties of H\&E stained images. This groundbreaking idea is rooted in the principles of Beer Lambert's law \cite{Rodger2013}, which relates to how objects absorb light. The proposed model showcases heightened stability concerning color inconsistency, making color normalization unnecessary in this context.

Furthermore, the architecture incorporates a Progressive Dense Feature Aggregation module, facilitating the effective merging and learning of features. The overall effectiveness of the proposed approach has been rigorously evaluated through ablation studies and robust experiments across three datasets, firmly establishing its efficacy in segmenting nuclei within whole slide images. The promising results demonstrated by the Triple U-Net further underscore its potential as a valuable tool for nuclei segmentation in diverse medical imaging applications.

\section{Dataset}
\label{sec:dataset}

Whole Slide Imaging (WSI) refers to the digital scanning of traditional glass slides containing cell tissues. This technique is widely used in digital pathology worldwide. The CryoNuSeg \cite{r1} dataset was created by selecting 30 WSIs from 10 different human organs, including the adrenal gland, larynx, lymph node, mediastinum, pancreas, pleura, skin, testis, thymus, and thyroid gland. These WSIs were obtained from The Cancer Genome Atlas \cite{tomczak2015cancer} and ensured data diversity by including samples from various scanning centers, different disease types, and both sexes. The images in the dataset have a magnification of 40x. Image patches of a fixed size of 512*512 pixels were extracted using QuPath \cite{bankhead2017qupath}. In order to carry out nuclei segmentation, two individuals with specialized training were involved: one was a biologist (referred to as Annotator 1), and the other was a bioinformatician (referred to as Annotator 2). Both annotators were provided with precise instructions for the manual marking of nuclei instances. Both annotators were given the exact instructions for segmentation. Annotator 1 performed a second round of manual mark-ups with a three-month gap to avoid intra-observer variability. Altogether, during the initial and subsequent rounds, Annotator 1 successfully identified and segmented 7,596 and 8,044 nuclei, respectively. In comparison, Annotator 2 performed the segmentation of 8,251 nuclei.

\section{Proposed Methodology}

\begin{figure}[htb]
  \centering
  \includegraphics[width=1\linewidth]{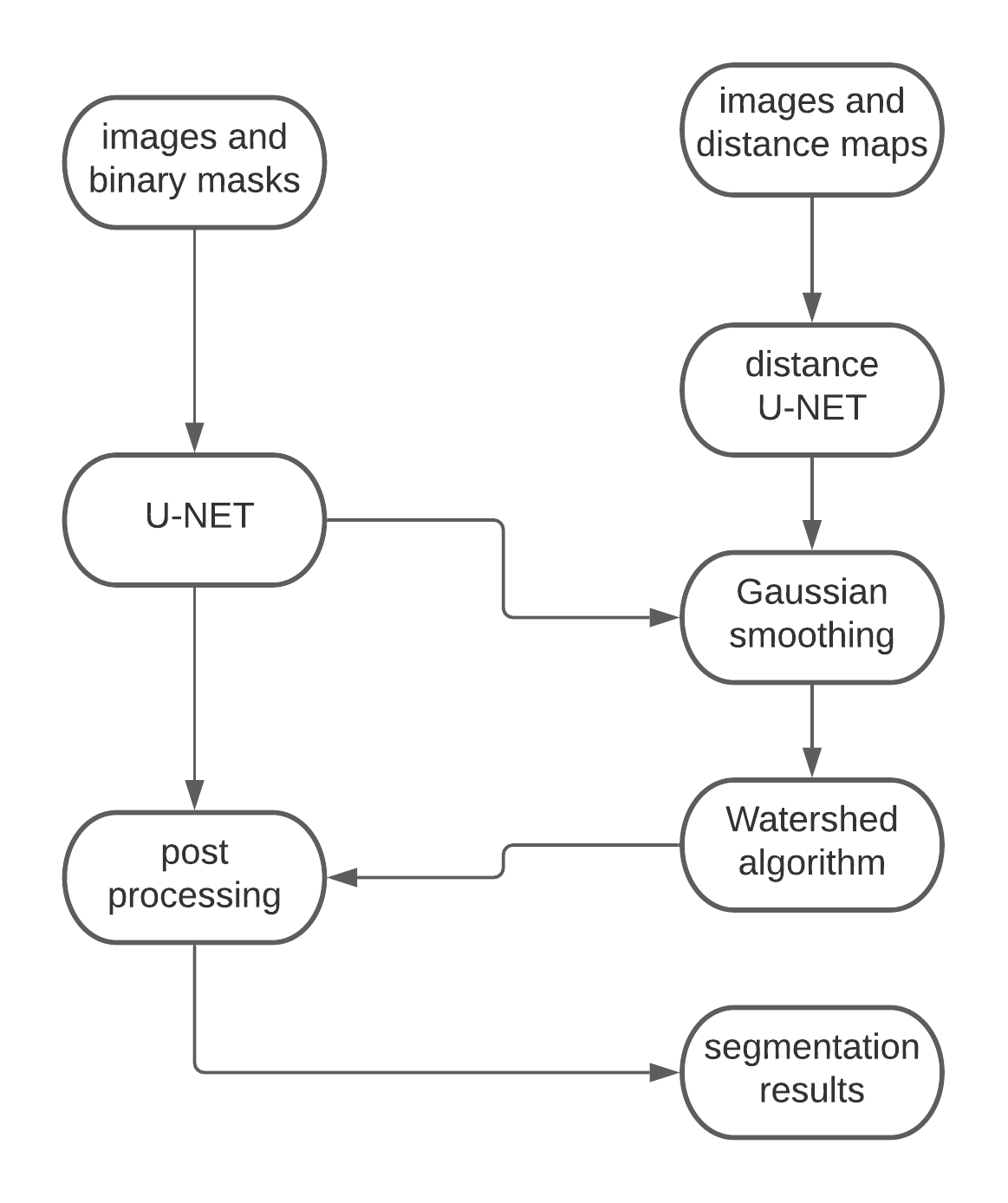}
  \caption{Baseline Algorithm used for Benchmark}
  \label{fig:base_arch}
\end{figure}

The authors introduced a U-NET based algorithm in the baseline paper \cite{r1} in which the dataset has been published. They used a recently published SOTA Deep Learning instance segmentation model as the baseline segmentation model. Below Figure \ref{fig:base_arch} shows a visual walkthrough of the whole baseline algorithm. The segmentation model consists of two sub-models, which are trained independently. A U-Net based model
(segmentation U-Net) is used to perform basic semantic segmentation, while a U-Net model focusing on the distance between centers of each nuclei is used to predict the distance maps for all nuclei instances. After training the two sub-models, their results are merged to form the instance segmentation masks.They first apply a Gaussian smoothing filter to obtain the final instance segmentation masks.Then, the local maxima are derived from the smoothed distance maps and are used as  points for a watershed algorithm. They use the segmentation U-Net results as the watershed method labels to determine the segmentation masks.

To achieve higher accuracy in nuclei segmentation, it is proposed to replace the traditional U-NET used in the baseline paper \cite{r1} with a better and improved version of U-NET. The proposed algorithm involves the utilization of Triple U-NET in place of U-NET, in addition to employing the watershed algorithm for post-processing and carrying out instance segmentation.

\begin{figure}[htb]
  \centering
  \includegraphics[width=1\linewidth]{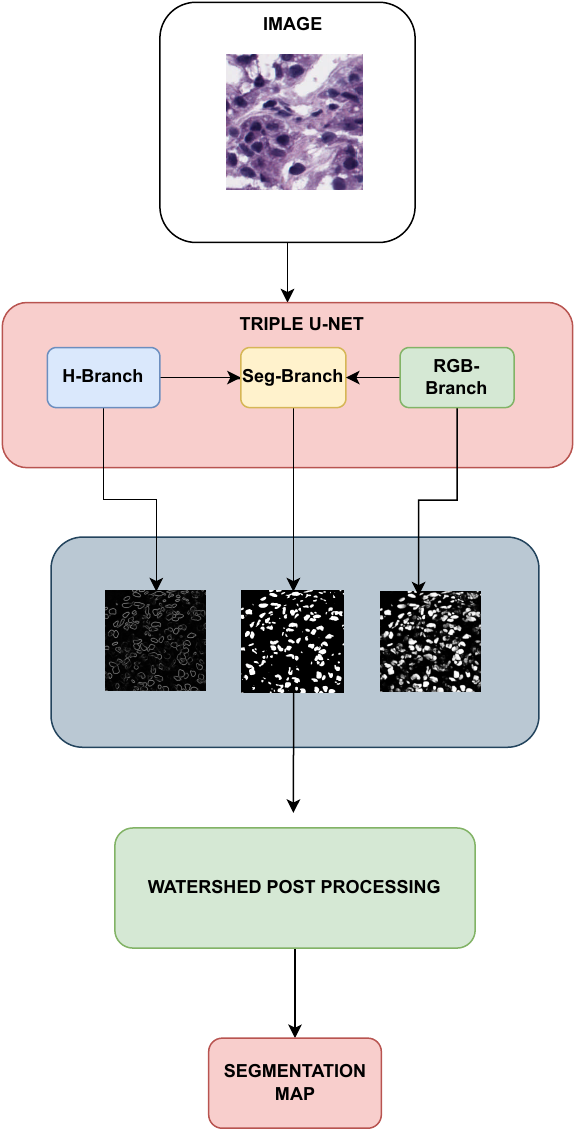}
  \caption{\textcolor{red}{Proposed Algorithm architecture}}
  \label{fig:proposed_arch}
\end{figure}

\begin{figure}[htb]
  \centering
  \includegraphics[width=1\linewidth]{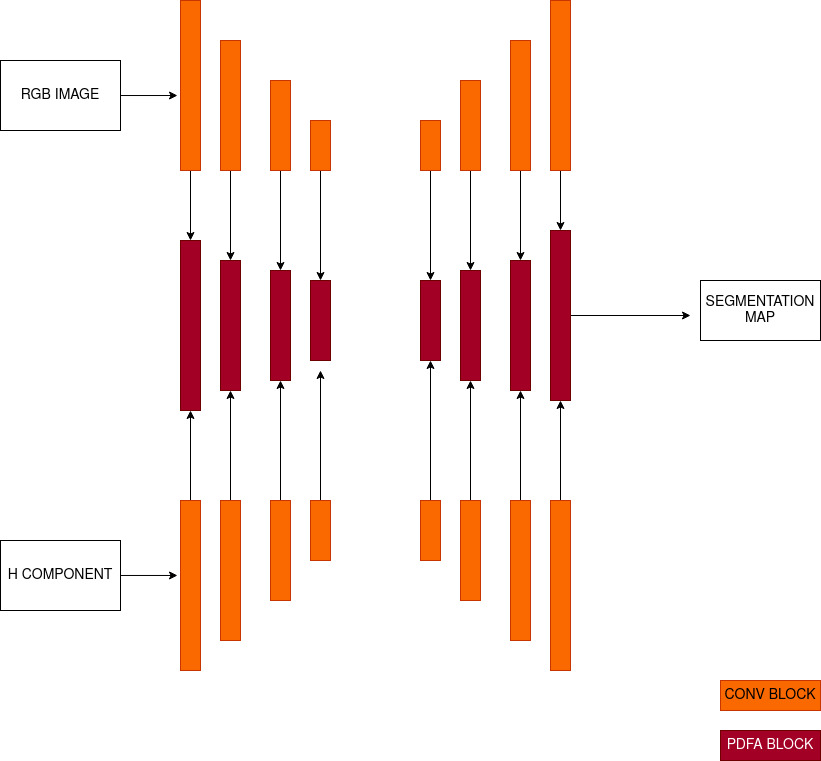}
  \caption{Triple U-Net architecture \cite{ZHAO2020101786}}
  \label{fig:triple_unet_architecture}
\end{figure}

The proposed architecture takes a cell tissue image as input, producing segmented areas of the nucleus and other cell parts as output. Figure \ref{fig:proposed_arch} provides a visual representation of the proposed architecture . Figure \ref{fig:triple_unet_architecture} provides a visual representation of the Triple U-Net model used in the proposed approach. The methodology has been elaborated in detail in this section, encompassing preprocessing, data augmentation, feature extraction, segmentation, evaluation metrics, and comparison.

A Triple U-Net based model \cite{zhao2020triple} was proposed for nuclear segmentation. It is basically a model with three u-net layers - the RGB branch, the H branch, and the segmentation branch. The RGB branch is for raw feature extraction, and the H branch is for contour detection tasks. The segmentation branch aggregates the results from the RGB and H branches and predicts the final results. A feature fusion strategy called Progressive Dense Feature Aggregation (PDFA) is introduced for better fusion and learning of feature representation.


\subsection{Dataset Preparation}

\subsubsection{Preprocessing}

The dataset consists of the RGB images and the corresponding binary masks. A binary mask is a grayscale image of the nucleus and the cell body. The nuclei are colored white, and the rest of the mask is black. The binary masks have been used to produce contour maps of the nuclei. This has been done by applying edge detection to the binary masks and identifying the boundaries of the nuclei in the image. The contour maps contain only the boundaries of the nuclei in white. Preprocessing has been done in this way because the Triple U-Net architecture has a branch called the H branch that works with the information of boundary detection from the contour map.

\subsubsection{Data Augmentation}
Deep learning, particularly in medical imaging, relies on a substantial amount of training data to achieve optimal performance and prevent overfitting. However, medical datasets often suffer from a limited number of images due to patient privacy concerns and the scarcity of specific tissue samples. To address this challenge, various techniques are used to create diverse variations of the same image, a process known as data augmentation. By generating new and distinct examples to train the datasets, data augmentation proves beneficial in enhancing the performance and outcomes of machine learning models. \\ 

A rich and sufficient dataset plays a crucial role in the improved and more accurate performance of machine learning models. In the field of medical image analysis, excessive data augmentation is commonly utilized by applying elastic deformations to the available training images. This approach enables the network to learn invariance to such deformations without having to witness these transformations in the annotated image corpus. This becomes particularly significant in biomedical segmentation, as deformation has been a prevailing variation in tissue and realistic deformations can be effectively simulated. The value of data augmentation for learning invariance has been
shown in \cite{NIPS2014_07563a3f} in the scope of unsupervised feature learning.
Different types of data augmentations have been used which include, random cropping, elastic transformation, rotation, flipping, and mirror. Each of the original images was cropped into 16 patches and fed into the neural network.

\subsection{Feature Extraction}
As discussed in the previous sections, the baseline algorithm used the U-Net model and the proposed algorithm uses the Triple U-Net model. One of the key differences between these two models is the Feature extraction method. The following sections discuss the feature extraction methods that is not present in the baseline U-Net model but are present in the Triple U-Net model and why it allows it to give much better results.
First, it will be discussed how hematoxylin extraction works. 
\subsubsection{Hematoxylin component extraction}

The technique of hematoxylin component extraction allows the differentiation of cytoplasmic, nuclei, and extracellular matrix features in an image. In the proposed method, the consistent staining properties of hematoxylin and eosin have been leveraged, where hematoxylin stains cell nuclei blue, and eosin stains the extracellular matrix and cytoplasm pink. To extract the Hematoxylin (H) component from the RGB color space, a color decomposition method has been applied based on Beer-Lambert's Law \cite{Rodger2013}. By isolating the H-component, a clearer view of the nuclei has been obtained. As a result, this significantly enhances the contrast between the nuclei and the cytoplasm as well as the stroma in the H\&E stained image. In this research paper, this characteristic has been utilized to assist in the segmentation task.
The basic U-Net does not focus on these properties.

\begin{figure}[htb]
  \centering
  \includegraphics[width=1\linewidth]{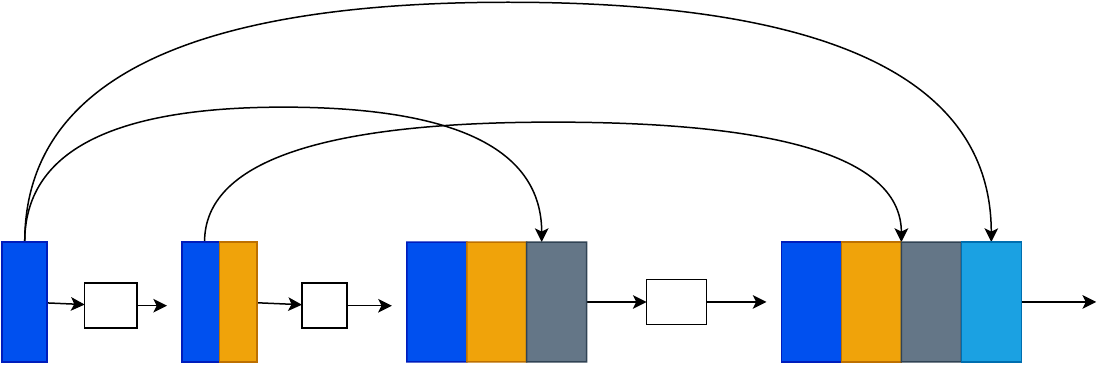} 
  \caption{PDFA Module \cite{ZHAO2020101786}}
  \label{fig:pdfa_architecture}
\end{figure}

\subsubsection{Progressive Dense Feature Aggression (PDFA) Module} 

In the implemented framework, distinct branches have been incorporated, namely RGB and H, to perform feature extraction for segmentation and nuclei contour detection tasks, respectively. Unlike conventional U-Net architectures that typically employ direct concatenation for fusing new features, a Progressive Dense Feature Aggression (PDFA) Module has been implemented, which progressively fuses new features. This unique feature fusion technique not only significantly reduces the number of parameters but also enhances feature propagation and reuse. The number of layers within a PDFA module varies depending on the features being fused; four layers have been used in the decoding phase and three layers in the encoding phase due to the absence of initial skipping features in the network.This feature concatenation technique is quite unique and is not present in the U-Net model.

\subsubsection{RGB branch}
The RGB branch extracts the foundational features needed for the segmentation task. Its designation as the "RGB branch" stems from its ability to extract raw data from the original image in the RGB color space, offering abundant semantic information crucial for segmentation. The outputs from this branch are supervised by comparing them to the segmentation ground truth through binary cross-entropy loss, a strategy that has proven exceptionally successful in acquiring the precise pixel-level features essential for segmentation. The loss function for the RGB branch is given below:

\begin{equation}
    \label{soft_dice_loss_for_rgb_branch}
     L_{RGB}=-1/N \sum_{i=1}^{N} (y_i * \log x_i+(1-y_i)*log(1-x_i)) 
\end{equation}
here, $x_i$ denotes the predicted probability of $i^{th}$ pixel, $y_i$ denotes the corresponding ground truth and N is the number of pixels.

\subsubsection{H branch}
The H branch gets the Hematoxylin component as its input. The Hematoxylin component is responsible for highlighting the nucleus's contour. This contour information is extracted from the Hematoxylin component and then used by the H branch to perform the task of contour detection. Utilizing this contour data, the model gains a deeper insight into the intricate contour particulars of the nucleus. As a result, the network attains enhanced accuracy in segmentation results, as the H branch helps establish more precise boundaries for the nucleus.
This contour awareness is also absent in default U-Net\\

To emphasize the learning of nuclear contour information, the Soft Dice Loss function is employed. The loss function is defined as follows:

\begin{equation} \label{soft_dice_loss_for_H_branch}
     L_H=1- \frac{2 \sum_{i=1}^{N}x_i y_i}{\sum_{i=1}^{N}x_{i}^{2} + \sum_{i=1}^{N}y_{i}^{2} }
\end{equation}

here, $x_i$ denotes the predicted probability of $i^{th}$ pixel and $y_i$ is the ground truth. 

\subsubsection{Segmentation Branch}
The segmentation branch assumes a pivotal role in amalgamating the raw attributes and the Hematoxylin-sensitive contour characteristics to produce the ultimate segmentation results. In this branch, a more sophisticated method for combining features, referred to as Progressive Dense Feature Aggregation (PDFA), is utilized to bolster the fusion process and facilitate feature representation learning across different branches. Rather than simply concatenating all the features directly, PDFA incrementally integrates preceding features with the subsequent ones.

\noindent

The newly introduced PDFA module offers a more coherent approach to feature fusion in contrast to the traditional densely connected block. It inherits the advantages associated with densely connected blocks, including the promotion of feature reuse, strengthening feature propagation, and a substantial reduction in the parameter count.

\noindent
The quantity of layers within an individual PDFA module is dictated by the quantity of new features that require fusion. As depicted in the diagram in figure \ref{fig:pdfa_architecture}, the PDFA module consists of four layers during the decoding phase, but it comprises three layers during the encoding phase due to the absence of initial skipping features at the network's beginning.

\noindent
 The Soft Dice Loss for the  segmentation branch is as follows:

\begin{equation} \label{soft_dice_loss_segmentation_branch}
     L_{segSD}=1- \frac{2 \sum_{i=1}^{N}x_i y_i}{\sum_{i=1}^{N}x_{i}^{2} + \sum_{i=1}^{N}y_{i}^{2} }
\end{equation}

\subsubsection{Loss Function}
Unlike the baseline algorithm, separate loss functions were used for each branch. The total loss function is the weighted summation of the losses from the three branches:

 \begin{equation}
     L_{total}= \lambda_{1}L_{RGB}+\lambda_{2}L_{H}+\lambda_{3}L_{seg\_ST}+\lambda_{4}L_{seg\_SD}  
 \end{equation}

The main benefits of Triple U-Net over the basic U-Net are as follows :

\begin{enumerate}
  \item No necessity of  color normalization 
  \item It generates segmentation results with highly accurate nucleus boundaries.
  \item The PDFA module is introduced to enable the network to progressively aggregate features from diverse domains.
\end{enumerate}

\subsection{Post Processing}

\subsubsection{Gaussian Filter}

Gaussian Filtering \cite{DHAEYER1989169} is an image noise reduction technique that uses the Gaussian Weighted function to smoothen images by convolving them with a Gaussian function. It functions as a filter that attenuates high-frequency elements within the images. Its equation is: \\

\begin{equation}
\label{gaussian_filter_1}
     P(x)=\frac{1}{\sqrt{2\pi\sigma^2}}e^{\frac{-x^2}{2\sigma^2}}
\end{equation}

\begin{equation}
\label{gaussian_filter_2}    
      P(x,y)=\frac{1}{\sqrt{2\pi\sigma^2}}e^{\frac{-x^2+y^2}{2\sigma^2}}
\end{equation}

Applying the equation \ref{gaussian_filter_2} to the images generates contours like concentric circles from the Gaussian distribution. The technique derives a convolution matrix and calculates weighted averages for each pixel, y removing high-frequency components and sharp edges. \\

\subsubsection{Watershed Algorithm}
\noindent
The Triple U-Net architecture produces segmentation maps from the input images. Marker-based image segmentation is performed using the watershed algorithm \cite{watershed} on these segmentation maps to achieve improved results. Watershed is a widely used image processing technique for segmentation, particularly important in differentiating overlapping regions, a critical aspect for precise instance segmentation. The algorithm needs user-defined markers which can be defined using thresholding techniques or morphological operations.  \\

The segmentation map can be considered as a topographic surface, where high-intensity regions denote peaks and low-intensity regions denote valleys. To separate different objects in the segmentation mask, which makes different nuclei in the segmentation map much clearer, the algorithm follows and simulates a flooding process of filling the valleys with water having different colors. The rising water from the isolated valleys will eventually rise to a level where they will start merging.  When this merging starts to occur, barriers have been built on top of the peak to prevent the peak from submerging. Once the barriers are built, they would make up the boundary of the object, and in this case, it is the boundary of the nucleus.  \\

\noindent

To prevent oversegmentation resulting from various noises that may be present in the segmentation map, the marker-based watershed algorithm \cite{6014611} is applied. In this process, the specification of all the valley points to be merged and those that should not be merged is carried out. This is accomplished by assigning distinct labels to known objects, indicating regions where the presence of nuclei is certain with different color values. A separate color is assigned to label the background, while the region about which certainty is uncertain is designated with the label 0. This is how the marker is created. Then watershed algorithm is applied and it produces a set of labels each of which corresponds to a different object. Each of the labels was then gone through, and the object was extracted. i.e. the different nuclei. This further improves the output produced by Triple U-Net to make the segmentation maps more prominent.

\subsection{Evaluation Metric}

Two types of evaluation metrics has been used - Average Jaccard Index (AJI) and Panoptic Quality (PQ). \\

\textbf{AJI}: AJI refers to the Average Jaccard Index. The Jaccard index is also known as the intersection over the union. The Jaccard similarity coefficient is a statistic for measuring the similarities and differences among samples. As the equation shows, Jaccard Similarity depends on two sets. It basically calculates how much overlap there is between the two sets. \\

% AJI = \frac{\sum_{i=1}^{n} $G_i$\bigcap_{}^{}$P_j$}{\sum_{i=1}^{n}$G_i$\bigcup_{}^{}$P_j$ + \sum_{k\in N}^{}$P_k$}  \\

\begin{equation}
\label{AJI}
  AJI = \frac {\sum_{i=1}^{n} G_i\bigcap P_j}
                {\sum_{i=1}^{n} G_i\bigcup P_j
                + \sum_{K\in N} P_k
                }
\end{equation}

In equation \ref{AJI}, G = { G 1 , G 2 , . . . , G n } and P = { P 1 , P 2 , . . . , P m } denote the ground truth and the prediction results respectively. N is the set of indices of prediction results without any corresponding ground truth. \\

\textbf{PQ}: The PQ (Panoptic Quality) is used to evaluate the performance of models in Panoptic Segmentation tasks. \\

% PQ = \frac{TP}{TP + 1/2*TP + 1/2*FN} * \frac{a}{b}\\

\begin{equation}
\label{pq}
  PQ =  \frac{ \left| TP \right| }
                {\left| TP \right| + 1/2\left| FP \right| + 1/2\left| FN \right| }
                \times
              \frac{\sum_{(x,y)\in TP}IoU(x,y)}
              {\left| TP \right|}
 \end{equation}

While both AJI and PQ measure instance segmentation performance, in the AJI score the correct classification in the numerator of the equation is divided by the sum of prediction results and ground truths. The correct classification is not being multiplied by 2. The effect of this is that the rate at which AJI penalizes wrong predictions is more drastic than it awards correct predictions. Therefore, AJI can be considered to be a more strict scoring method and is a more challenging metric to achieve high scores with. On the other hand, PQ tends to reward correct predictions more. The AJI score in the implementation outperforms the baseline model by a significant margin.

\subsection{Implementation and Experiment Training Details}

\noindent The entire experiment was conducted utilizing Google COLAB as the computational platform. The network has been implemented using Pytorch Framework with a GPU of Nvidia Tesla K80. \\\\
Throughout the training process, the input images were augmented using various transformation techniques, such as elastic transformation, random cropping, mirror rotation, and flipping. These augmentations were applied to the original images, which were initially resized to 256x256 pixels. The batch size was fixed to 10. \\\\
During the training procedure, the learning rate was varied from 0.001 to 0.002. To achieve this, various learning rate schedulers were tested, such as Exponential LR reduction on Plateau, Cosine annealing, and Cosine annealing restarts.

\label{sec:methodology}

\section{Results and Discussion}
To present and validate the obtained results cross validation has been performed on the dataset.The individual organ's images are divided into folds and a 10-fold cross-validation is performed. This increases the reliability of our obtained results

\label{sec:results}
\subsection{Segmentation Results}
Robust evaluation procedures have been performed to show the reliability of the achieved results. The same evaluation procedure which has been used by the authors who provided the baseline has been followed. To ensure comparability with the baseline, the exact dataset division method used by the authors is followed to obtain the results. During the experiments, 10-fold cross-validation has been conducted on the Cryonuseg dataset. The total dataset comprises 30 images of 10 organs. There are 3 images per organ and each fold consists of 3 images with one image per organ. The Triple-Unet model has been used for this experiment where training is done using 27 images and 3 images are kept as a hold-out set. The model performance is tested on the hold-out set where each time the hold-out set contains 3 images of a different organ. In the experiments, comparable or improved results are obtained across all three metrics for each organ.(PQ, AJI, DICE). The average score overall 10 folds are AJI score of 0.6741416 \%, and PQ score of 50.56\%.The results are shown in Table-1

\begin{table}
    \centering
    \begin{tabular}{|l|l|l|l|}
    \hline

        Model & AJI Score & PQ Score \\ \hline

       U-Net & 52.5 & 47.7  \\ \hline
       Triple U-Net & \textbf{67.41} & \textbf{50.56}  \\ \hline
    
    \end{tabular}
    \caption{Summary of Performance Comparison	}
\end{table}

\begin{figure}
  \centering
  \includegraphics[width=0.9\linewidth]{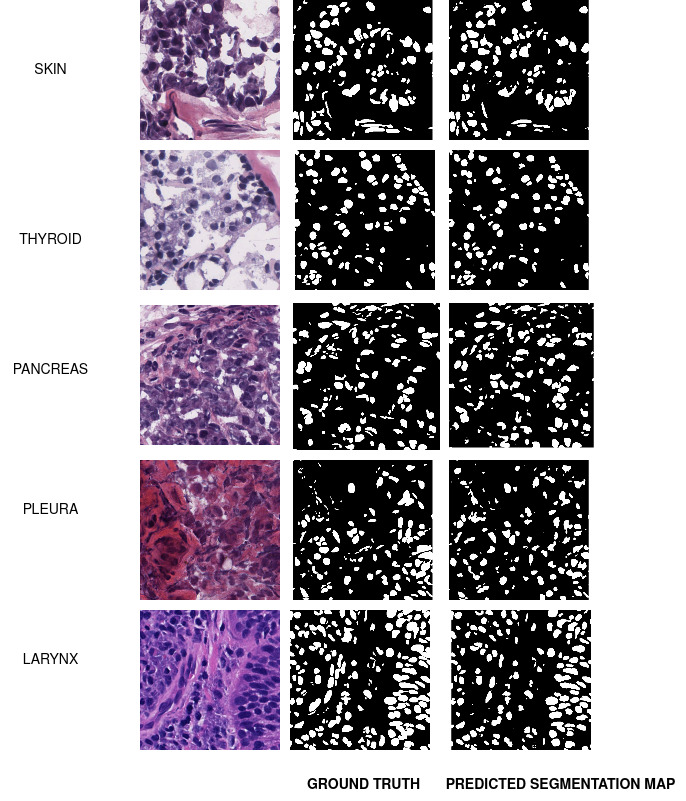} % Replace "path/to/your/image.png" with the actual path to your image file.
  \caption{Segmentation results}
  \label{fig:tissue_images}
\end{figure}

\begin{table*}
\centering 
\begin{tabular}{|l|ll|ll|}
\hline
\multicolumn{1}{|c|}{\multirow{2}{*}{Organ}} & \multicolumn{2}{c|}{AJI Score (\%)}                  & \multicolumn{2}{c|}{PQ Score (\%)}                        \\ \cline{2-5} 
\multicolumn{1}{|c|}{}                       & \multicolumn{1}{l|}{Baseline Model} & Our Model      & \multicolumn{1}{l|}{Baseline Model}      & Our Model      \\ \hline
\textbf{adrenal gland}                       & \multicolumn{1}{l|}{53.49}          & 66.82          & \multicolumn{1}{l|}{48.30}               & 55.03          \\ \hline
\textbf{larynx}                              & \multicolumn{1}{l|}{59.70}          & 72.04          & \multicolumn{1}{l|}{54.50}               & 54.37          \\ \hline
\textbf{lymph node}                          & \multicolumn{1}{l|}{53.54}          & 72.66          & \multicolumn{1}{l|}{50.79}               & 49.94          \\ \hline
\textbf{mediastinum}                         & \multicolumn{1}{l|}{54.10}          & 63.90          & \multicolumn{1}{l|}{50.73}               & 48.08          \\ \hline
\textbf{pancreas}                            & \multicolumn{1}{l|}{44.84}          & 63.74          & \multicolumn{1}{l|}{37.75}               & 48.02          \\ \hline
\textbf{pleura}                              & \multicolumn{1}{l|}{46.49}          & 62.71          & \multicolumn{1}{l|}{40.02}               & 45.72          \\ \hline
\textbf{skin}                                & \multicolumn{1}{l|}{47.84}          & 70.72          & \multicolumn{1}{l|}{40.78}               & 51.27          \\ \hline
\textbf{testis}                              & \multicolumn{1}{l|}{50.49}          & 68.40          & \multicolumn{1}{l|}{47.51}               & 47.76          \\ \hline
\textbf{thymus}                              & \multicolumn{1}{l|}{56.46}          & 65.79          & \multicolumn{1}{l|}{52.83}               & 48.07          \\ \hline
\textbf{thyroid gland}                       & \multicolumn{1}{l|}{58.20}          & 67.36          & \multicolumn{1}{l|}{53.48}               & 57.29          \\ \hline
\textbf{Average}                             & \multicolumn{1}{l|}{\textbf{52.5}}  & \textbf{67.41} & \multicolumn{1}{l|}{\textbf{47.7}} & \textbf{50.56} \\ \hline
\end{tabular}
 \caption{Segmentation Results and Organ Wise Comparison with Baseline}
\end{table*}

% \usepackage{multirow}

\subsection{Comparison with Baseline}

Comparison of the proposed model performance with the baseline model is shown in Tables 2, 3, and 4. 
% \usepackage{multirow}
% \usepackage{float}

\begin{table}[H]
\begin{tabular}{ |p{2cm}|p{2cm}|p{2cm}|p{2cm}|  }
 \hline
 \multicolumn{3}{|c|}{AJI Score} \\
 \hline
Fold & Baseline & Our Model\\ \hline
% 0  & 0.7817 & 0.796866   \\ \hline
        0  & 0.5349 & 0.668186   \\ \hline
        1  & 0.5970 & 0.720366   \\ \hline
        2  & 0.5354 & 0.72661   \\ \hline
        3  & 0.5410 & 0.639007   \\ \hline
        4  & 0.4484 & 0.637411   \\ \hline
        5  & 0.4649 & 0.62711   \\ \hline
        6  & 0.4784 & 0.707169   \\ \hline
        7  & 0.5049 & 0.684007   \\ \hline
        8  & 0.5646 & 0.657913   \\ \hline
        9 & 0.582 & 0.673637   \\ \hline
        \textbf{AVERAGE} & \textbf{0.525} & \textbf{0.6741416}  \\ \hline
\end{tabular}
\caption{Comparison of AJI Score}
\end{table}

\begin{table}[H]
\begin{tabular}{ |p{2cm}|p{2cm}|p{2cm}|p{2cm}|  }
 \hline
 \multicolumn{3}{|c|}{PQ Score} \\
 \hline
Fold & Baseline & Our Model\\ \hline
% 0  & 0.550355  & 0.4830  \\ \hline
        0  & 0.4830 & 0.550355   \\ \hline
        1  & 0.5450 & 0.543708   \\ \hline
        2  & 0.5079 & 0.499412   \\ \hline
        3  & 0.5073 & 0.480857   \\ \hline
        4  & 0.3775 & 0.480179   \\ \hline
        5  & 0.4002 & 0.457219   \\ \hline
        6  & 0.4078 & 0.512735   \\ \hline
        7  & 0.4751 & 0.477618   \\ \hline
        8  & 0.5283 & 0.4807   \\ \hline
        9 & 0.5348 & 0.572905  \\ \hline
        \textbf{AVERAGE} & \textbf{0.477} & \textbf{0.5055688}  \\ \hline
\end{tabular}
\caption{ Comparison of PQ Score}
\end{table}

\subsection {Comparison with other U-Net Variants}
The results have been compared with other U-Net variants such as Nested U-Net with Efficient Encoder \cite{nest}. The algorithm with the Triple U-Net performs better than its counterparts as shown in Table 5.

\begin{table}[H]
\captionsetup{justification=centering}
\begin{tabular}{|p{4cm}|p{1cm}|p{1cm}|}
\hline
\textbf{Model} & \textbf{AJI} & \textbf{PQ} \\
\hline
Triple Unet with Watershed Post-processing & 0.674 & 0.506 \\

\hline
Baseline Unet with Watershed Post-processing & 0.525 & 0.477 \\
\hline
Nested Unet with EfficientNet Encoder & 0.604 & 0.476 \\
\hline

\end{tabular}
 \caption{Segmentation Results Compari- \\\hspace{0.5cm}    son with Other U-net Variants}
\end{table}

\subsection{Performance Analysis}

Our proposed architecture surpasses the baseline model across all evaluation metrics. Notably, it exhibits the highest improvement in the AJI metric, achieving an impressive nearly 30 percent increase in score from 52.5 to 67.41.
The Triple U-net achieves edge detection with remarkable accuracy and produces more precise contour-aware boundaries. The PDFA allows the network to learn features and merge their information from various domains of the nuclei. An additional advantage of the proposed  model is that it does not require any form of color normalization, resulting in significant savings in computational power.\\
In conclusion, the proposed architecture stands as a state-of-the-art solution for high-precision nuclei instance segmentation of cryosectioned tissues.

\subsection{Error Analysis}

The model's performance was comparatively lower for Mediastinum, Larynx, and Lymph Node samples due to limited training data, complex tissue cells, and smaller nuclei that are difficult to detect. The scarcity of cancer cases in these organs further hampers data availability for training. Additionally, the image quality of frozen samples worsens segmentation. To improve results, a larger and diverse dataset with more samples from these organs is necessary. Augmenting the training data with examples from these specific organs is expected to enhance the model's performance in nuclei instance segmentation for Mediastinum, Larynx, and Lymph Node samples.

\section{Conclusion}
\label{sec:conclusion}

It is proposed to implement a state-of-the-art algorithm including the Triple-Unet model as the core part, for precise nuclei instance segmentation in cryosectioned H\&E stained frozen tissues. Triple U-Net focuses on properties specific to H\&E stained images that are widely used in the medical domain and produce much better results than a general U-Net model. This will help to analyze nuclei morphology, size, and density enabling accurate abnormality detection. Consequently, it aids in rapid cancer diagnosis, potentially saving lives. Moreover, the solution empowers surgeons to make confident intraoperative decisions by providing essential information. In conclusion, if implemented as a real-life solution, this cutting-edge approach has the potential to provide invaluable assistance in comprehensive tissue analysis, early cancer diagnosis, and informed surgical interventions.


% % use section* for acknowledgment
% \ifCLASSOPTIONcompsoc
%   % The Computer Society usually uses the plural form
%   \section*{Acknowledgments}
% \else
%   % regular IEEE prefers the singular form
% \fi




%\bibliographystyle{IEEEtran}


% \end{thebibliography}

\bibliographystyle{IEEEtran}
\bibliography{bibliography/references}

\end{document}